\documentclass{webofc}
\usepackage[varg]{txfonts}   
\usepackage{tikz} 
\usetikzlibrary{arrows} 
\usetikzlibrary{shapes,arrows,positioning,automata,backgrounds,calc,er,patterns}
\usepackage{tikz-feynman}
\usepackage{hyperref}

\newcommand{\op}{\mathcal{O}}
\newcommand{\mF}{\mathcal{F}}

\newcommand{\mO}{\mathcal{O}}
\begin{document}
\title{SMEFT as a slice of HEFT's parameter space}
\author{\firstname{Alexandre} \lastname{Salas-Bern\'ardez}\inst{1} \and
        \firstname{Juan J.} \lastname{Sanz-Cillero}\inst{1} \and
        \firstname{Felipe J.} \lastname{Llanes-Estrada}\inst{1} \and
        \firstname{Raquel} \lastname{G\'omez-Ambrosio}\inst{2}
}

\institute{Universidad Complutense de Madrid, Dept. F\'{\i}sica Te\'orica and IPARCOS, 
Fac. CC. Fisicas, Plaza de las Ciencias 1, 28040 Madrid Spain.
\and
Dipartimento di Fisica ``G. Occhialini”, Universit\`a degli Studi di Milano-Bicocca,
and INFN, Sezione di Milano Bicocca, Piazza della Scienza 3, I – 20126 Milano, Italy
          }

\abstract{%
The Standard Model Effective Field Theory (SMEFT) is the parametrization chosen to interpret many modern measurements. We have recently discussed, building on the work of other groups, that its overall framework can be experimentally tested, beyond simply constraining its parameters. This is because the Higgs Effective Field Theory (HEFT) is somewhat more general, as it does not assume that the Higgs boson $h$ needs to be embedded in 
a complex doublet $H$ on which the Standard Model (SM) and SMEFT are built. 
As a result, the HEFT parameter spaces for the various relevant channels contains hypersurfaces over which one may use  SMEFT to describe data. If experimental measurements of HEFT's parameters in any of those various channels yield a point outside of any of the hypersurfaces, SMEFT is falsified; meanwhile, its framework remains appropriate (in particular, as long as the SM remains compatible with data).
A common necessity of the various possible tests is that processes involving different number of Higgs bosons (maintaining the number and nature of other particles unchanged) need to be contrasted. 
}
\maketitle
%

\section{Introduction: two EFTs for the electroweak sector}
\label{sec-1}

Beyond the Standard Model (BSM) searches in experiments at the LHC and other machines have come to be used to constrain
the well-known SMEFT expansion,
\begin{equation}
    \mathcal{L}_{\rm SMEFT} =
    \mathcal{L}_{\rm SM} + 
    \sum_{n=5}^{\infty}
    \sum_i
    \frac{c_i^{(n)}}{\Lambda^{n-4}} \op_i^{(n)}(H) \ .
 \label{SMEFTL}
 \end{equation}
There, the various operators are organized by canonical (field and derivative) dimension as exposed by the powers of $\Lambda$ made explicit, and their $c_i$ coefficients bound by experiment~\cite{Grinstein:2007iv,Grzadkowski:2010es,Li:2020gnx,Ellis:2020unq}.  The particle content of the electroweak symmetry breaking sector is packaged in a Higgs doublet field, just as in the SM,
\begin{equation}
H = 
\frac{1}{\sqrt{2}} \begin{pmatrix} \varphi_1+i\varphi_2 \\   \varphi_0 + i\varphi_3
\end{pmatrix} 
= 
U(\boldsymbol{\omega}) \begin{pmatrix} 0 \\   (v+h_{\rm SMEFT})/\sqrt{2}
\end{pmatrix} \, .
\end{equation} 
There, the Cartesian coordinates $\varphi_a$ can be rearranged into the polar decomposition in terms of the $\omega_i$ Goldstone bosons (which set the orientation of $H$ through the unitary matrix $U(\boldsymbol{\omega})$) and the radial coordinate $h_{\rm SMEFT}$ (with $|H|=(v+h_{\rm SMEFT})/\sqrt{2}$). If, alternatively,
no assumption about the Higgs-boson $h$ being part of a doublet $H$ is made, the appropriate effective theory is rather HEFT, directly written in terms of $h$ (see~\cite{Dobado:2019fxe} for a recent review)
\begin{align}  \label{HEFTL}
{\cal L}_{\rm HEFT} = \frac{1}{2}\partial_\mu h_{\rm HEFT}\partial^\mu h_{\rm HEFT}-V(h_{\rm HEFT}) +\frac{1}{2}\!\mathcal{F}(h_{\rm HEFT})
\partial_\mu \omega^i \partial^\mu \omega^j\!\left(\!\delta_{ij}\!+\!\frac{\omega^i\!\omega^j}{v^2\!-\!\boldsymbol{\omega}^2}\!\right)\ .
\end{align}
The $h_{\rm HEFT}$ Higgs (or for simplicity, $h$) is exposed in multiplicative functions rescaling the traditional operators of the old electroweak chiral Lagrangian; for an alternative form see~\cite{Graf:2022rco}.

It was believed that one could indistinctly convert them into one another, as between Cartesian and polar coordinates, 
by wrapping or unwrapping the $H$ field to expose $h$ and the Goldstone bosons $\omega_i$, or to reconstruct it from them. 
But several works led by the San Diego group~\cite{Alonso:2016oah,Alonso:2016btr,Alonso:2021rac,Cohen:2020xca} have clarified that while HEFT can always be cast as a SMEFT, the converse is not always true, there are some requirements on the ``flare'' function of the Higgs field 
$
    {\mathcal F}(h_{\rm HEFT})=1+\sum_{n=1}^{\infty}{a_n}\Big(\frac{h_{\rm HEFT}}{v}\Big)^n \,,
$
that we will shortly expose; saliently, $\mathcal{F}$ needs to have a zero $h^\ast$ and satisfy certain analyticity properties. 

In our recent contributions~\cite{Gomez-Ambrosio:2022qsi,Gomez-Ambrosio:2022why}, we have translated these criteria into practical conditions that amount to finding certain correlations among the HEFT parameters that, if violated by experimental data (also interpretable in terms of HEFT coefficients~\cite{Buchalla:2015qju,Delgado:2013hxa}), would falsify SMEFT. Here we quickly overview these developments.

\section{Regularity conditions for the applicability of SMEFT}

Alonso, Jenkins and Manohar~\cite{Alonso:2015fsp} formulated HEFT in geometrical terms in the space of Higgs-fields and detailed the regularity conditions that the HEFT Lagrangian needs to satisfy to support a SMEFT, employing an elegant covariant formalism. 
We will proceed in a more pedestrian way, converting SMEFT into HEFT and viceversa with   
 \begin{align}
& |\partial H|^2 +  
\frac{1}{2} B(|H|^2)(\partial (|H|^2))^2 \longleftrightarrow \frac{v^2}{4} \mathcal{F}(h_{\rm HEFT}) \,   {\rm Tr}\{ \partial_\mu U^\dagger \partial^\mu U\} 
+
\frac{1}{2} (\partial h_{\rm HEFT})^2 \,,
\end{align}  
The change from SMEFT to HEFT is straightforward and always possible, with the canonical, nonlinear change of variables given in differential form as
\begin{equation}  \label{changeofvariable}  
dh_{\rm HEFT}\, =\, \sqrt{1+(v+h_{\rm SMEFT})^2 B(h_{\rm SMEFT})}\,\, dh_{\rm SMEFT} \,,
\end{equation}
where the flare-function is provided by $ \mathcal{F}(h_{\rm HEFT}) \,=\, \left(1+h_{\rm SMEFT}/v\right)^2$.  

However,  the reciprocal conversion 
$ h_{\rm HEFT}   \,=\, \mathcal{F}^{-1}\left((1+h_{\rm SMEFT}/v)^2\right)$ from HEFT to SMEFT is not always possible, 
because of the need to reconstruct squared operators of the Higgs doublet field $H$ necessary for SMEFT, such as
\begin{align}
|H|^2 &= \frac{(v+h_{\rm SMEFT})^2}{2}\, ,
\nonumber \\
(\partial|H|^2)^2 &= (v+h_{\rm SMEFT})^2 \,   (\partial h_{\rm SMEFT})^2 \nonumber =\, 2 |H|^2 \,   
(\partial h_{\rm SMEFT})^2 \, .
\end{align} 
The extra $|H|^2$ on the right hand side of the second equation 
appears then in a denominator
\begin{align}
&\mathcal{L}_{\rm HEFT} =  \underbrace{ |\partial H|^2}_{= \mathcal{L}_{\rm SM}} \quad +\quad 
\label{eq:HEFT2SMEFT}\underbrace{ \frac{1}{2} \bigg[  \frac{8|H|^2}{v^2}\bigg(   (\mathcal{F}^{-1})'\left(2| H |^2/v^2\right)  \bigg)^2 \,\,\,-\,\,\, 1\bigg] \, \frac{(\partial| H |^2)^2}{2| H |^2}  }_{=\Delta \mathcal{L}_{\rm BSM}}\, .
\end{align}
As SMEFT requires a Taylor expansion as in Eq.~(\ref{SMEFTL}), this singularity precludes its existence and needs to be cancelled by the preceding bracket in the second line of Eq.~(\ref{eq:HEFT2SMEFT}).

The existence of that zero $h^\ast\equiv \mathcal{F}^{-1}(0)$ of $\mathcal{F}$, and the analyticity required for a power series expansion for $\mathcal{F}$, the Higgs potential $V$, etc., become necessary requirements for a HEFT Lagrangian density to be expressible as a SMEFT.
Our findings, in agreement with~\cite{Cohen:2020xca}, can be abstracted as follows.
\begin{enumerate}
    \item{} The $\mathcal{F}$ function in the HEFT Lagrangian density of Eq.~(\ref{HEFTL}) $\mathcal{F}(h^\ast)=0$ must have a double zero;  $SU(2)\times SU(2)$ is a good global symmetry there.
    
    \item{} There, its second derivative must satisfy
    $ \mathcal{F}''(h^\ast)=\frac{2}{v^2}$.
        
    \item{} 
 All odd derivatives vanish, $\mathcal{F}^{(2\ell+1)}(h^\ast)=0$.  
\end{enumerate}

The restrictions over $\mathcal{F}$ at the symmetric zero $h^\ast$ ($h=-v$ in the SM) that guarantee the existence of a SMEFT 
are theoretically illuminating but not too practical, since our measurements are taken around the vacuum $h=0$.

\section{Translating the geometrical conditions into constraints over the accessible HEFT parameter space}

\begin{figure}[hb!]
  \begin{minipage}[c]{0.45\textwidth}
    \includegraphics[width=\textwidth]{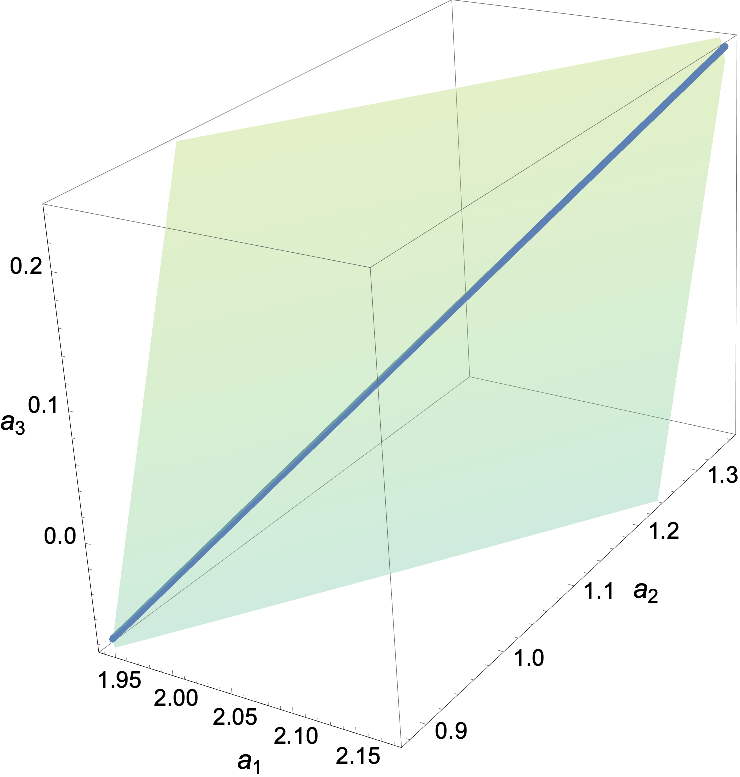}
  \end{minipage}\hfill
  \begin{minipage}[c]{0.45\textwidth}
    \caption{ Correlations for the HEFT coefficients from Table~\ref{tab:correlations} that need to be satisfied for SMEFT to be a valid EFT of the electroweak symmetry breaking sector. The solid diagonal is the correlation  of order $\Lambda^{-2}$, that becomes broadened as the indicated band (greenish online) at order $\Lambda^{-4}$.}   \label{fig:aicorrelations}
    \end{minipage}
\end{figure}

A key contribution of our earlier work has been to translate those conditions into restrictions over the $a_i$ at the physical vacuum $h=0$. To do it, we match two series expansions, the one around $h=0$ (setting $a_0:=1$ and taking $h$ normalized to $v$, so that $v=1$)
and the one around $h^\ast$, expanding in terms of $a^\ast_j=\mF^{(j)}(h^\ast)/j!$,
\begin{equation}
     \mathcal{F}(h) = \sum_{i=0}^n a_i h^i\ , \ \ \ \ \ \mathcal{F}(h) = \sum_{j=0}^n a^\ast_j (h-h^\ast)^j\ .
\end{equation}
By matching the two expansions around the two different points it is easy to read off the coefficients $a^\ast_j$ that encode the conditions over $\mathcal{F}$ in terms of the $a_i$, accessible to experiment. The resulting correlations among the HEFT conditions are given in Table~\ref{tab:correlations}. The relation between the three lowest-order coefficients involving one, two and three $h$ bosons, (respectively $a_1$, $a_2$, $a_3$)
is shown in Figure~\ref{fig:aicorrelations}. Finally, the numerical intervals for these HEFT parameters $a_i$ currently consistent with experimental data are given 
in Table~\ref{tab:correlations2}.

\begin{table}[hb!]
    \caption{\small Correlations among the $a_i$ HEFT coefficients necessary for SMEFT to exist, at the orders $\Lambda^{-2}$ and $\Lambda^{-4}$, given in terms of $\Delta a_1:=a_1-2=2a-2$ and $\Delta a_2:=a_2-1=b-1$ (so that all entries in the table vanish in the SM, with all equalities becoming $0=0$). 
    Notice that the r.h.s. of each identity in the second column shows the $\mO(\Lambda^{-4})$ corrections to the relations of the first column. The third one assumes the perturbativity of the SMEFT expansion.
    \label{tab:correlations}}
    \centering
    \begin{tabular}{|c|c|c|}\hline
        \textbf{Correlations}  & \textbf{Correlations}  & 
       ${\Lambda^{-4}}$ \textbf{Assuming}  
        \\ 
        \textbf{accurate at order} $\Lambda^{-2}$ & \textbf{accurate at order} $\Lambda^{-4}$ & \textbf{SMEFT perturbativity}
        \\ \hline
        $\Delta a_2=2\Delta a_1$ &  &   $|\Delta a_2| \leq 5 |\Delta a_1|$
        \\
        $a_3=\frac{4}{3} \Delta a_1$ & 
         $\left(a_3-\frac{4}{3}\Delta a_1\right) =\frac{8}{3}(\Delta a_2-2 \Delta a_1)   -\frac{1}{3}\left(\Delta a_1\right)^2$ 
        & 
        \\  
        $a_4=\frac{1}{3} \Delta a_1$ & 
        $\left(a_4-\frac{1}{3}\Delta a_1\right) = \frac{5}{3}\Delta a_1 - 2\Delta a_2 +\frac{7}{4} a_3=  $
        & 
        those for $a_3$, $a_4$, $a_5$, $a_6$
        \\
         & $ \phantom{\left(a_4-\frac{1}{3}\Delta a_1\right)}=\frac{8}{3}(\Delta a_2-2 \Delta a_1)-\frac{7}{12}\left(\Delta a_1\right)^2  $ &
        \\ 
        $a_5=0$ &  
        $a_5 = \frac{8}{5}\Delta a_1 -\frac{22}{ 15} \Delta a_2 +a_3=$ 
        & 
        all the same 
        \\ 
         &  $\phantom{a_5}= \frac{6}{5}  (\Delta a_2-2 \Delta a_1) -\frac{1}{3}\left(\Delta a_1\right)^2  $   & 
        \\ 
       $a_6=0$ & 
        $a_6=\frac{1}{6}a_5$
        &
         \\ 
        \hline
    \end{tabular}
\end{table}

\begin{table}[hb!]  
\setlength{\arrayrulewidth}{0.3mm} 
\setlength{\tabcolsep}{0.2cm}  
\renewcommand{\arraystretch}{1.4} 
    \caption{\small Bounds on the $a_i$ HEFT coefficients that need to be satisfied for SMEFT to exist. The first column being consistent at $\mathcal{O}(\Lambda^{-2})$ with 95\% confidence-level~\cite{ATLAS:2020qdt}) experimental bounds $a_1/2 \in [0.97,1.09]$.
    The second column corresponds to the next order~\cite{Gomez-Ambrosio:2022qsi}. 
    \label{tab:correlations2}}
    \centering
    \begin{tabular}{|c|c|}\hline
         $\mathcal{O}(1/\Lambda^2)$ & $\mathcal{O}(1/\Lambda^4)$ \\ \hline 
         $\Delta a_2\in [-0.12,0.36] $  &  \\
          $a_3\in[-0.08,0.24]$  & $a_3\in [-3.1,1.7]$  \\  
         $a_4\in [-0.02,0.06]    $ &  $a_4\in [-3.3,1.5]$ \\
         & $a_5 \in[-1.5,0.6]$ \\
           & $ a_6=a_5$ \\ 
        \hline
    \end{tabular}
\end{table}

\begin{figure}[!ht]
  \begin{minipage}[c]{0.67\textwidth}
    \includegraphics[width=\textwidth]{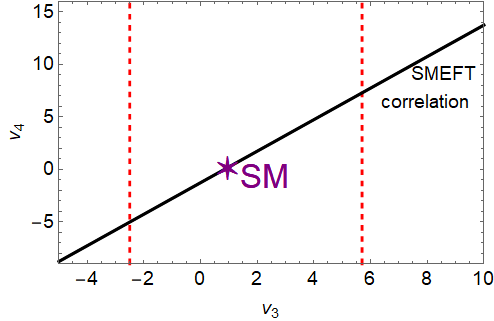}
  \end{minipage}\hfill
  \begin{minipage}[c]{0.3\textwidth}
    \caption{
\small The correlation $v_4=\frac{3}{2} v_3 -\frac{5}{4} -\frac{1}{6}\Delta a_1$ that SMEFT predicts at $\mathcal{O}(1/\Lambda^2)$, from the 95\% confidence interval for $v_3\in[-2.5,5.7]$~\cite{ATLAS:2021jki}. The experimental $a_1$ uncertainty~\cite{ATLAS:2020qdt,CMS:2022cpr},  $\frac{a_1}{2}\in[0.97,1.09]$, is numerically negligible and allows to predict the (solid, diagonal) SMEFT band. An experimental measurement of $v_4$ is necessary to test this, by adding a horizontal band to the plot.
    } \label{fig:a1a22}
  \end{minipage}
\end{figure}

We have next examined a couple of other HEFT multiplicative form-factor functions 
that incorporate powers of $h$ to further operators of the electroweak Lagrangian; our second example here is the nonderivative $V(H)$ Higgs-potential, accessible at ``low'' $\sqrt{s}$ as the absence of derivatives minimizes the number of momentum powers.
The traditional SM form
\begin{equation}
{\mathcal{L}}_{\rm SM} = |\partial H|^2 -   \underbrace{\left( \mu^2 |H|^2 +\lambda|H|^4\right)}_{V(H)} \, ,
\end{equation}
written in terms of $h$ acquires additional non-renormalizable couplings in HEFT, organized in a power-series expansion
\begin{align}\label{expandV}
    V_{\rm HEFT}= \frac{m_h^2 v^2}{2}  \Bigg[&   \left(\frac{h_{\rm HEFT}}{v}\right)^2 +  v_3   \left(\frac{h_{\rm HEFT}}{v}\right)^3 + v_4    \left(\frac{h_{\rm HEFT}}{v}\right)^4 + \dots \Bigg]\,,
\end{align}
with $v_3=1$, $v_4=1/4$ and $v_{n\geq 5}=0$ in the SM. 
For the applicability of SMEFT, the coefficients of this series must satisfy the constraints listed in Table~\ref{tab:corV} and Figure~\ref{fig:a1a22}.
From the ATLAS~\cite{ATLAS:2021jki} bound $\Delta v_3\in[-2.5,5.7]$, $\mO(1/\Lambda^2)$ SMEFT predicts the coefficient intervals in the last column of Table~\ref{tab:corV}, which can be tested in few-Higgs final states. A coupling 
$c_{H\Box}\neq 0$ (stemming from the SMEFT operator relevant at the TeV scale for the electroweak symmetry breaking sector $\mathcal{O}_{H\Box}=|H|^2\Box|H|^2$ \cite{Gomez-Ambrosio:2022qsi}) introduces the correction $\Delta a_1\propto c_{H\Box}$ that appears in the first row (and affects the flare function already discussed), but in the potential is for now numerically negligible since $v_3$ experimental uncertainties are way larger than the allowed $\Delta a_1$.

\begin{table}[!b]
\setlength{\arrayrulewidth}{0.3mm} 
\setlength{\tabcolsep}{0.2cm}  
\renewcommand{\arraystretch}{1.4} 
\caption{\small First two rows: correlations among the coefficients $\Delta v_3:=v_3-1$, $\Delta v_4:=v_4-1/4$, $v_5$ and $v_6$ of the HEFT Higgs potential expansion in Eq.~(\ref{expandV}) that need to hold, at $\mO(1/\Lambda^2)$, if SMEFT is a valid description of the electroweak sector.
The third row includes the leading correlations for the Yukawa $\mathcal{G}(h)$ function of Eq.~(\ref{Yukawa}), constraining $c_2$ and $c_3$ by $c_1$ and $a_1$ (from the correction to the value of the symmetric point $h^*$). We make use of current 95\% confidence interval for the top Yukawa coupling $c_1\in[0.84,1.22]$~\cite{deBlas:2018tjm}.}
\label{tab:corV}
\begin{center}
 \begin{tabular}{|c|c|} \hline 
        $\Delta v_4=\frac{3}{2}\Delta v_3  -\frac{1}{6}\Delta a_1$ & $\Delta v_4\in[-3.8,8.6]$\\[2ex]
        $ v_5=6v_6=\frac{3}{4}\Delta v_3 -\frac{1}{8}\Delta a_1 $ 
        & $v_5=6 v_6 \in[-1.9,4.3]$ \\ \hline 
        $c_2
       = 3 c_3  
        =\frac{3}{2} (c_1 -1) -\frac{1}{4}\Delta a_1$ & $c_2
        = 3 c_3  
        \in[-0.27,0.35]$\\
         \hline    
    \end{tabular}
\end{center}  
\end{table}

\begin{figure}
  \begin{minipage}[c]{0.67\textwidth}
    \includegraphics[width=\textwidth]{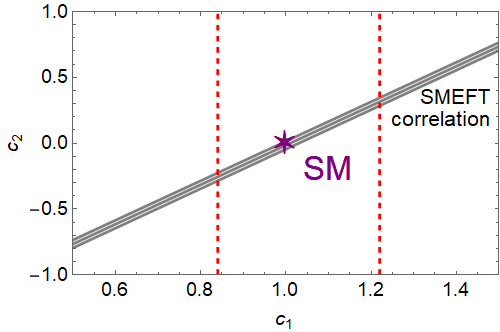}
  \end{minipage}\hfill
  \begin{minipage}[c]{0.3\textwidth}
    \caption{
The correlation $c_2=\frac{3}{2} (c_1 -1) -\frac{1}{4}\Delta a_1$ predicted by SMEFT at $\mathcal{O}(1/\Lambda^2)$ in the top Yukawa sector of HEFT.
The 95\% confidence interval for the coupling $c_1\in[0.84,1.22]$ is taken from~\cite{deBlas:2018tjm} (red dashed lines). The experimental  uncertainty~\cite{ATLAS:2020qdt,CMS:2022cpr} on $a_1$, namely  $a_1/2\in[0.97,1.09]$ (at 95\% CL), is reflected in the width of the gray band with the SMEFT correlation. 
    } \label{fig:c1c2}
  \end{minipage}
\end{figure}

As a third relevant example, we proceed from the SM  coupling the top quark to the Higgs boson. This has been extended into HEFT~\cite{Castillo:2016erh} by a multiplicative function $\mathcal{G}(h)$ 
\begin{equation}
\mathcal{L}_Y= -\mathcal{G}(h) M_t \bar{t} t 
\sqrt{1-\frac{\boldsymbol{\omega}^2}{v^2}}\; ,
\end{equation}
that can be Taylor-expanded around the physical $h=0$ vacuum according to
\begin{equation} \label{Yukawa}
    \mathcal{G}(h_{\rm HEFT})= 1 + c_1 \frac{h_{\rm HEFT}}{v} + c_2 \left( \frac{h_{\rm HEFT}}{v} \right)^2+\dots \end{equation}  
    (in the SM $c_1=1$ and $c_{i\geq 2}= 0$ simplify).
The correlations among these coefficients induced by SMEFT at order $1/\Lambda^2$  are also given, in the last row of Table~\ref{tab:corV} and in Figure~\ref{fig:c1c2}.

\newpage

\section{Conclusion: \\
Testing the embedding of SMEFT into HEFT at colliders}

The SM is compatible, to date, with the measurements of the electroweak sector. It is a particular case of SMEFT and as a matter of course, all correlations discussed in this article among the coefficients of HEFT (many trivially) are satisfied.

Nevertheless, if the high-luminosity LHC of a future accelerator such as a new $e^-e^+ $ machine finds cracks in the SM, testing~\cite{Gonzalez-Lopez:2020lpd} whether both SMEFT and HEFT or only HEFT can be applied will become an issue.

The $\omega\omega\to nh$ processes with $n$ a varying number of Higgs particles would be key to disentangling the EWSBS. They give direct access to the $a_i$ coefficients of the flare function $\mathcal{F}$, and hence to their correlations, as listed in Table~\ref{tab:correlations}.
Figure~\ref{fig:Feynman} shows the typical Feynman diagrams describing the contact part of these processes that yield direct access to the $a_i$ coefficients.

\begin{figure}[!ht]  
\centering
     \begin{tikzpicture}[scale=.7]
     \draw[decoration={aspect=0, segment length=1.8mm, amplitude=0.7mm,coil},decorate] (-1,1) -- (0,0.)-- (-1,-1);
          \draw[] (1.35,0.0) node {$h$};
     \draw[] (0,0)-- (1.1,0);
       \draw[] (2.75,0) node {$\displaystyle{  \, =\,  -\frac{a_1}{2v} \,  s }$};  
            \draw[] (1.25,1.2) node {$\;$};
          \draw[] (1.25,-1.2) node {$\;$};
\end{tikzpicture}\;\;
     \begin{tikzpicture}[scale=.7]
     \draw[decoration={aspect=0, segment length=1.8mm, amplitude=0.7mm,coil},decorate] (-1,1) -- (0,0)-- (-1,-1);
     \draw[] (0,0)-- (1,1);
     \draw[] (1.25,1) node {$h_1$};
          \draw[] (1.25,-1) node {$h_2$};
     \draw[] (0,0)-- (1,-1);
       \draw[] (2.75,0) node {$\displaystyle{  \, =\,  -\frac{a_2}{v^2} \,  s }$};  
\end{tikzpicture}\;\;
     \begin{tikzpicture}[scale=.7]
     \draw[decoration={aspect=0, segment length=1.8mm, amplitude=0.7mm,coil},decorate] (-1,1) -- (0,0)-- (-1,-1);
     \draw[] (0,0)-- (1,1);
     \draw[] (0,0)-- (1.1,0);
     \draw[] (1.25,1) node {$h_1$};
     \draw[] (1.35,0.0) node {$h_2$};
          \draw[] (1.25,-1) node {$h_3$};
     \draw[] (0,0)-- (1,-1);
       \draw[] (2.75,0) node {$\displaystyle{  \, =\,  -\frac{3!a_3}{2v^3} \,  s }$};  
\end{tikzpicture}
\caption{\label{fig:Feynman}\small
Contact diagrams which provide $a_i$; in a given physical process they are summed to $t$-channel exchanges between the two initial Goldstone bosons (wiggly lines) $\omega_i$, as given in~\cite{Gomez-Ambrosio:2022qsi}.
}
\end{figure}

For example, the BSM part of the two- to two-body amplitude is given by a single SMEFT coefficient,
\begin{equation}
    T_{\omega\omega \to hh}^{\rm SMEFT}(s)=-{2s}\frac{c_{H\square}}{\Lambda^2}+\mathcal{O}(\Lambda^{-4})\ .
\end{equation}
At this same order, the two- to three-body one receives a contribution from the $a_3$ coefficient and also $t$-channel exchanges; in HEFT, these are controlled by $a_1$ and $a_2$, and are independent. 
However, in SMEFT all three coefficients are functions of the same $c_{H\square}$ and appear to cancel, $T_{\omega\omega \to hhh}^{\rm SMEFT}(s)=0+\mathcal{O}(\Lambda^{-4}).$

\begin{figure}[ht!]
\begin{minipage}[c]{0.6\textwidth}
\includegraphics[width=\textwidth]{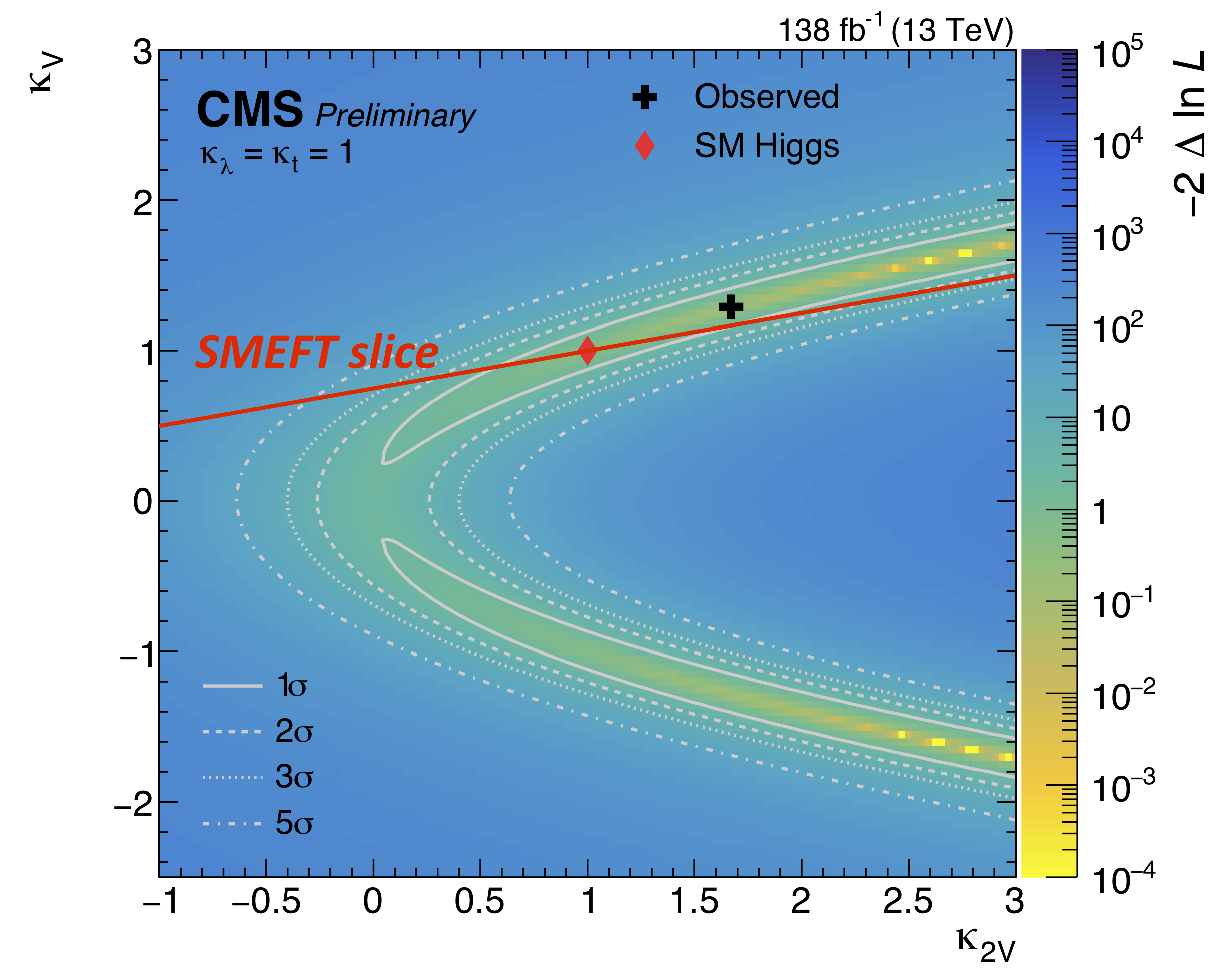}
\end{minipage}
\begin{minipage}{0.38\textwidth}
\caption{ Experimental CMS exclusion plot
in the $a_1$, $a_2$ plane
(reproduced from~\cite{CMS:2022nmn} under the 
\href{https://creativecommons.org/licenses/by/4.0/legalcode}{Creative Commons BY 4.0 license}.)
We have superimposed the SMEFT slice through this parameter space, that is seen to run parallel to one of the branches of the experimental exclusion band. 
\label{fig:withdata}}
\end{minipage}
\end{figure}

Figure~\ref{fig:withdata} displays the currently allowed experimental band (highlighted in yellow) for the first two coefficients, $a_1$ and $a_2$ (in the popular $\kappa$ notation in which the experimental cross section is scaled to the SM one, these are $\kappa_V$ and $\kappa_{2V}$)~\footnote{We would like to thank Nick Smith for pointing out the experimental result for $\kappa_V,\ \kappa_{2V}$ in~\cite{CMS:2022nmn}.}.

We have superimposed the SMEFT line (red online) which represents the $\mathcal{O}(1/\Lambda^2)$ correlation between these two parameters.  

This shows that current data is much more sensitive to $a_1$ than $a_2$, but the gentle sloping almost parallel to the SMEFT slice through that HEFT parameter space can also signal that  the measurement is more sensitive to the 
separation from SMEFT than to the coefficient of SMEFT itself. This might be due to the fact that $c_{H\square}$ only brings an overall renormalization of $h$ and no additional derivatives, so it does not change the shape of any kinematic distributions, it is a parameter of more difficult access.
The SM (and therefore SMEFT too) is seen to comfortably sit within the $1-\sigma$ band, in good agreement with the data.

Current experimental constraints are not really significant for  $n\geq 2$ 
and it will be challenging to obtain information on the coefficients for higher number of Higgs bosons $n$.
Better reconstruction techniques at the high-luminosity LHC run but especially a future high-energy collider (either hadronic or muonic) will hopefully improve the situation.

Likewise, to exploit the correlation in Figure~\ref{fig:c1c2}, an experimental determination of the 
$ t\bar{t}\to hh$ coupling $c_2$ needs to be undertaken.   If it would be measured, 
SMEFT can be tested by comparing to $c_2\in[-0.27,0.35]$  (where uncertainties have been added linearly). Similar measurements in the Yukawa sector, with an increasing number of Higgs bosons, would lead to further tests of the SMEFT framework.
Similar considerations apply to the Higgs potential $V(h)$.

To conclude, while other authors have also proposed methods to try to distinguish SMEFT from HEFT~\cite{Cohen:2021ucp},
we believe that our work offers clear criteria that depend only on the availability of sufficiently precise experimental data, and eventually higher order EFT computations~\cite{Hays:2018zze} to match that precision.

\newpage
\bibliography{references}

\section*{Acknowledgments}
Supported by Spanish PID2019-108655GB-I00/AEI/10.13039/501100011033 grant, Universidad Complutense de Madrid under research group 910309 and the IPARCOS institute; 
ERC Starting Grant REINVENT-714788; UCM CT42/18-CT43/18; the Fondazione Cariplo and Regione Lombardia, grant 2017-2070.

\end{document}